\documentclass[aip,graphicx]{revtex4-1}

\usepackage{algorithm, amsmath, amsthm, amssymb, amsfonts, amsbsy}
\usepackage[titletoc,page]{appendix}
\usepackage{bm, booktabs}
\usepackage{caption, color, colortbl}
\usepackage{enumerate}

\usepackage{latexsym, graphicx, epsfig, eqnarray}
\usepackage{natbib, version}
\usepackage{mathtools, multirow, mathrsfs}
\usepackage[mathscr]{eucal}

\usepackage{subfigure}
\usepackage[english]{babel}
\usepackage{float}
\usepackage{systeme}
\usepackage[noend]{algpseudocode}

\usepackage{}
\usepackage{pifont}
\usepackage{tabularx}
\usepackage{tikz,mathpazo}
\usepackage{url}
\usepackage{verbatim} 
\usepackage{graphicx}
\usepackage{hyperref}

\draft 

\begin{document}


\title{PCNN: a physics-constrained neural network for multiphase flows \footnote{This article may be downloaded for personal use only. Any other use requires prior permission of the author and AIP Publishing. This article appeared in Haoyang Zheng, Ziyang Huang, Guang Lin, A physics-constrained neural network for multiphase flows, Phys. Fluids 34, 102102 (2022) and may be found at \hyperlink{https://aip.scitation.org/doi/full/10.1063/5.0111275}{https://doi.org/10.1063/5.0111275}.}} 



\author{Haoyang Zheng}
\email[]{zheng528@purdue.edu}

\affiliation{School of Mechanical Engineering, Purdue University, West Lafayette, IN 47907, USA}

\author{Ziyang Huang}
\email[]{ziyangh@umich.edu}

\affiliation{Department of Mechanical Engineering, University of Michigan, Ann Arbor, MI 48109, USA}

\author{Guang Lin}
\email[]{guanglin@purdue.edu}
\thanks{Corresponding author}
\affiliation{School of Mechanical Engineering, Purdue University, West Lafayette, IN 47907, USA}
\affiliation{Department of Mathematics, Purdue University, West Lafayette, IN 47907, USA}


\begin{abstract}
The present study develops a physics-constrained neural network (PCNN) to predict sequential patterns and motions of multiphase flows (MPFs), which includes strong interactions among various fluid phases.
To predict the order parameters, which locate individual phases in the future time, a neural network (NN) is applied to quickly infer the dynamics of the phases by encoding observations. The multiphase consistent and conservative boundedness mapping algorithm (MCBOM) is next implemented to correct the predicted order parameters. This enforces the predicted order parameters to strictly satisfy the mass conservation, the summation of the volume fractions of the phases to be unity, the consistency of reduction, and the boundedness of the order parameters. Then, the density of the fluid mixture is updated from the corrected order parameters. Finally, the velocity in the future time is predicted by another NN with the same network structure, but the conservation of momentum is included in the loss function to shrink the parameter space. The proposed PCNN for MPFs sequentially performs (NN)-(MCBOM)-(NN), which avoids nonphysical behaviors of the order parameters, accelerates the convergence, and requires fewer data to make predictions. Numerical experiments demonstrate that the proposed PCNN is capable of predicting MPFs effectively.
\end{abstract}

\pacs{}

\maketitle 

\section{Introduction}

\subsection{Literature review}

Multiphase flows (MPFs) include motions of various phases and deformations of interfaces formed by the phases. Interactions among the phases are highly non-linear and complex. Although it is possible to predict MPFs by collecting their historical information, such as the flow rate, temperature, pressure, and positions of the phases, the mechanism is so complex that it is still difficult to build a general mathematical model to describe different kinds of MPFs or to solve those models efficiently and accurately. Predicting MPFs becomes a popular topic and many researchers have focused on it because they benefit a wide range of fields, e.g., the chemical industry, environmental protection, life sciences, and industrial process \cite{abolhasani2016oscillatory,yue2018multiphase,behari2020chronic,gidaspow2018kinetic,huang2021consistentNM,huang2021consistentSolid}.

In recent years, the development of neural network (NN) technology \textcolor{black}{\cite{dissanayake1994neural, lagaris1998artificial}} promotes time series modeling, which also contributes to predicting MPFs. With the use of self-organizing NNs, Mi \emph{et al.} \cite{mi2001flow} simulated the vertical bubbly, slug, churn, and annular flows for both the impedance of the multiphase admixture and the flow patterns. Based on the given experimental and theoretical velocity data, Valero and Bung \cite{valero2018artificial} identified the characteristic shapes of the fluid phases and explored the behaviors of the air-water flow. Rashid \emph{et al.} \cite{rashid2019prediction} developed a radial basis function NN to predict MPFs under critical conditions and discovered the relationships between the upstream pressure and gas-liquid ratio, and between the choke bin size and liquid flow rate. Serra \emph{et al.} \cite{serra2020two} proposed a randomized hough transform with an NN to predict the void fraction with given image samples, and their results are consistent with the actual void fraction values in natural circulation-based systems of nuclear power plants. Xuan \emph{et al.} \cite{xuan2022data} proposed a data-driven NN model to predict the separation-induced transition. Lino \emph{et al.} \cite{lino2022multi} proposed a graph neural network to extrapolate the time evolution of the unsteady Eulerian fluid dynamics. More works on MPFs simulation with the help of NNs were summarized in Yan$^{\prime}$s \cite{yan2018application} and \textcolor{black}{Karniadakis$^{\prime}$s \cite{karniadakis2021physics}} reviews.

To model MPFs, parameters inside NN will be updated until flow characteristics predicted by NN are similar to or even the same as the given samples. Some popular optimization algorithms \cite{al2018virtual, mo2019deep, kingma2014adam} can be applied to update the parameters inside NN. However, how to incorporate physical laws when optimizing the parameter space of NNs so that the predictions strictly meet physical constraints is still a problem. Consequently, NN can easily produce nonphysical results, especially when the number of samples is insufficient to train a complex NN with a large parameter space. Thus, recent studies are considering including physical constraints to improve the performance of NNs in the sense that not only available data are more effectively used, but also the burden of data acquisition is alleviated. The physical constraints can be enforced either implicitly or explicitly. For implicit constraints, penalties are added to the loss function and one obtains the so-called physics-informed neural networks (PINNs) \cite{raissi2019physics}. For explicit constraints, predictions from NN are modified to satisfy the physical constraints exactly \cite{huang2020consistentCH,huang2020consistentCAC,huang2020consistentN,huang2021consistentVolume}.

Physics-informed neural networks (PINNs) aim to include the physical constraints implied in the available data, by adding penalties in the loss function \cite{raissi2019physics}. This helps to reduce the hypothesis space of parameters, and thus requires fewer observations. Moreover, an NN model with desired physical properties is more acceptable and explainable in scientific problems because it has a better generalizable property in out-of-sample scenarios \cite{willard2020integrating}. Pang \emph{et al.} \cite{pang2019fpinns} developed fractional physics-informed neural networks (fPINNs) to encode fractional-order partial differential equations (PDEs). Yang and Perdikaris \cite{yang2019adversarial} proposed probabilistic PINNs to approximate arbitrary conditional probability densities and to generate samples similar to the one given by PDEs. \textcolor{black}{Dwivedi and Srinivasan \cite{dwivedi2020physics} considered the advantages of PINNs and extreme learning machine, and proposed Physics Informed Extreme Learning Machine (PIELM) to speed up the network training process. Leake and Mortari \cite{leake2020deep} transformed PDEs into unconstrained problems, and then solved them by training NNs.} Goswami \emph{et al.} \cite{goswami2020transfer} proposed a PINN algorithm to solve brittle fracture problems and optimize its loss by decreasing the variational energy of the system. \textcolor{black}{Schiassi \emph{et al.} \cite{schiassi2021extreme} proposed a novel PINNs framework to satisfy DE constraints analytically but with unconstrained parameters in NNs, which was also applied to thermal creep flow \cite{de2021physics} and Poiseuille flow simulations \cite{de2022physics}.} Qiu \emph{et al.} \cite{qiu2022physics} proposed PINNs for incompressible phase field method, which can capture the dynamic behavior precisely. Many existing studies have indicated that an NN with physical constraints contributes to improving prediction accuracy \cite{tartakovsky2018learning, meng2020ppinn,karpatne2017physics, jia2019physics, read2019process}, discovering PDEs and governing equations \cite{geneva2020modeling, zhu2019physics}, modeling inverse problems \cite{raissi2019deep, kahana2020obstacle}, and solving uncertainty quantification \cite{yang2019highly, yang2019adversarial}. Some other studies also focus on considering a general deep learning framework to learn diverse continuous nonlinear operators \cite{lu2021learning, li2020neural, li2020fourier}. Therefore, adding penalties from the physical constraints in the loss function becomes more and more acceptable and interpretable when developing NNs for physical problems.

\subsection{Our contribution: novelties and outline}
In the present study, we proposed a PCNN to predict the temporal dynamics of MPFs. Our method partly remedies a defect of PINNs that the physical constraints are not enforced exactly. Specifically to MPFs, the defect of PINNs may produce local voids, overfilling, mass loss, negative volume fractions, or negative density of the fluid mixture. In the present method, three physical constraints \cite{huang2021consistentVolume} on the order parameters (OPs), which locate individual phases, are exactly satisfied by correcting the prediction from the NNs, instead of adding penalties to the loss function. The first constraint is called the summation constraint, which requires the summation of the volume fractions to be unity. The second one is called the conservation constraint, which requires the mass (or volume) of each phase to be conserved if there are no external inputs. The last one is called the boundedness constraint, which requires the volume fractions produced by OPs to be between zero and one. We demonstrate in the present study that the nonphysical behaviors due to failures of satisfying the three physical constraints will be propagated into the training process and gradually accumulated. As a result, large errors are introduced as time goes on. Once we obtain the prediction of OPs, we proceed to compute the density of the fluid mixture, which later becomes the input to the velocity prediction with the constraint of momentum conservation.

The structure of the proposed PCNN for MPFs is as follows. The OPs at the new time level are first predicted from the NN model with the OPs at previous time steps. The time-dependent dynamics of the OPs are encoded by a recurrent neural network and decoded predicted OPs for the next time step. The predicted OPs are then corrected by the multiphase consistent and conservative boundedness mapping (MCBOM) algorithm \cite{huang2021consistentVolume} to strictly satisfy the four physical constraints mentioned, and the corrected OPs are the final prediction of the OPs at the new time level. As a result, one can update the density of the fluid mixture using the OPs at the new time level. Finally, the velocity at the new time level is predicted from the velocity at previous time levels and the density of the fluid mixture, using the NN model where the momentum conservation is included as a penalty in the loss function. Considering the momentum conservation as a penalty can shrink the model parameter space towards initial momentum with respect to the maximum likelihood estimates.

The remaining sections are structured as follows. In Section \ref{Sec_Problem_definition}, the problem of interest is defined. Section~\ref{proposed} introduces the proposed PCNN for MPFs. Then, in Section~\ref{experiment}, we verify the feasibility and practicability of the proposed model through two applications. Finally, conclusions are drawn in Section~\ref{conclusion}.

\section{Problem definition}\label{Sec_Problem_definition}

In this section, we define the MPFs to be modeled, as well as their physical constraints and analyze the MPFs with time series.

\subsection{MPFs}

In the present study, the incompressible MPFs are considered, where there are $N$ $(N \geqslant 1)$ immiscible fluid phases. Each phase has a fixed density and viscosity, denoted by $\rho^p$ and $\mu^p$, respectively, for phase $p$, and every two-phase has a surface tension, denoted as $\sigma^{p,q}$ for phases $p$ and $q$. Locations of the phases are determined from the order parameters $\{\phi^p\}_{p=1}^N$, such that $\phi^p=1$ represents pure phase $p$ while phase $p$ is absent at $\phi^p=-1$.

Three physical constraints need to be strictly satisfied by the order parameters. The first one is called the summation constraint, i.e.,
\begin{equation}\label{Eq Sum phi}
\sum_{p=1}^N \frac{1+\phi^p}{2}=1
\quad
\mathrm{or}
\quad
\sum_{p=1}^N \phi^p=2-N.
\end{equation}
The summation constraint requires that the summation of the volume fractions of all the phases is one so that local voids or overfilling is not produced. 

The second constraint is the mass (or volume) conservation, i.e.,
\begin{equation}\label{Eq Conservation phi}
\frac{d}{dt} \int_{\Omega} \phi^p d\Omega
+
\int_{\partial \Omega} (\mathbf{n} \cdot \mathbf{v}) \phi^p d\Gamma
=0,
\quad
1 \leqslant p \leqslant  N,
\end{equation}
where $\Omega$ is the domain of interest, $\partial \Omega$ is the domain boundary, $\mathbf{n}$ is the unit outward normal at the domain boundary, and $\mathbf{v}$ is the velocity. If there is no normal velocity at the domain boundary, the integral of $\phi^p$ $(1 \leqslant  p \leqslant  N)$ over the domain of interest does not change. As a result, the volume and mass of each phase, i.e., $\int_{\Omega} \frac{1+\phi^p}{2}d\Omega$ and $\int_{\Omega} \rho^p \frac{1+\phi^p}{2}d\Omega$, respectively, $1 \leqslant  p \leqslant  N$, do not change with time either.

The third constraint is the boundedness constraint, which requires
\begin{equation}
    |\phi^p| \leq 1, \quad \forall t>0,
    \quad
    1 \leq p \leq N.
\end{equation}
One can physically explain $\{\phi^p\}_{p=1}^N$ as volume fraction contrasts only when they are in $[-1,1]$. This corresponds to the volume fractions $\{\frac{1+\phi^p}{2}\}_{p=1}^N$ in $[0,1]$. 
\textcolor{black}{
Here, the volume fraction contrast of Phase~$p$ is defined as $\phi_p=C_p-\sum_{q=1,q \neq p}^N C_q$, where $\{C_p\}_{p=1}^N$ is the volume fractions of the phases.}

The density of the fluid mixture is
\begin{equation}\label{Eq Density}
    \rho=\sum_{p=1}^N \rho^p \frac{1+\phi^p}{2},
\end{equation}
From Eq. (\ref{Eq Density}), if $\phi^p$ is outside $[-1,1]$, the density of the fluid mixture can become negative. Then, $(\rho \mathbf{v})$ is the momentum and should be conserved if there are no external forces at the domain boundary.

\subsection{Time series prediction in MPFs}

Time series is a sequence of data set arranged in chronological order \cite{hamilton1994time, weber2001visualizing}. The time series tool is applied to predict the dynamics of MPFs, or specifically, the order parameters and flow velocities of MPFs. Given the time series of the observed order parameters $\{\phi_i\}_{i=1}^{m}$ and flow velocity $\{v_i\}_{i=1}^{m}$, the goal is to predict the unobserved data in the future, i.e., $\{\phi_i\}_{i=m+1}^n$ and $\{v_i\}_{i=m+1}^n$. Here, $m$ and $n$ are the numbers of the observed time steps and total (including both observed and unobserved) time steps, respectively. The subscripts $i$ is timestamp indexes for the data. To achieve the goal, we specify two projections, where one $f_1(\cdot)$ from $\boldsymbol\phi_k$ to $\boldsymbol\phi_{k+1}$, and another $f_2(\cdot)$ from $\boldsymbol v_k$ to $\boldsymbol v_{k+1}$ such that:
\begin{equation}\label{function1}
\boldsymbol\phi_{i+1}=f_1(\boldsymbol\phi_i)+\boldsymbol\varepsilon_1,\ \ \ \ \boldsymbol v_{i+1}=f_2(\boldsymbol v_i)+\boldsymbol\varepsilon_2,
\end{equation}
where  $\boldsymbol\phi_i=\{\phi_i\}_{i}^{t+i}$, $\boldsymbol\phi_{i+1}=\{\phi_{i+1}\}_{i}^{t+i}$, $\boldsymbol v_i = \{v_i\}_{i}^{t+i}$, and $\boldsymbol v_{i+1} = \{v_{i+1}\}_{i}^{t+i}$ $(1 \leqslant i \leqslant n-t-1)$. Here $\boldsymbol\varepsilon_1$ and $\boldsymbol\varepsilon_2$ are both bias vectors $\{\varepsilon_i\}_{i}^{t+i}$ ($\varepsilon_i\sim \mathcal{N}(0, 1)$). $t$ $(1 \leqslant t \leqslant n)$ is the sequence length.
To approximation $f(\cdot)$, we consider neural networks $\widehat{f}(\cdot)$:
\begin{equation}\label{estimations}
\widehat{\boldsymbol\phi}_{i+1}=\widehat{f}_1(\boldsymbol\phi_i, \boldsymbol\theta_1),\ \ \ \ \widehat{\boldsymbol v}_{i+1}=\widehat{f}_2(\boldsymbol v_i, \boldsymbol\theta_2),
\end{equation}
where $\boldsymbol \theta_1^*=\arg\min\limits_{\boldsymbol \theta_1}\|f_1(\boldsymbol \phi)-\widehat{f}_1(\boldsymbol\phi, \boldsymbol\theta_1)\|$ and $\boldsymbol \theta_2^*=\arg\min\limits_{\boldsymbol \theta_2}\|f_2(\boldsymbol v)-\widehat{f}_2(\boldsymbol v, \boldsymbol\theta_2)\|$ are parameters to minimize the difference between ground truth and prediction. In the next section, we will elaborate the way to design our neural networks and to incorporate the physical laws inherent in MPFs.

\section{The proposed model to predict the multiphase flow}\label{proposed}

This section will elaborate on the proposed PCNN for multiphase flow prediction, including the NN model to predict the order parameters in Section \ref{cnp_lstm_method}, the boundedness mapping (MCBOM) from Huang \emph{et al.} \citep{huang2021consistentVolume} to correct the order parameters in Section \ref{conceive}, and the second NN model to predict the velocity in Section \ref{pinns}.

\subsection{The neural network to predict the order parameters}\label{cnp_lstm_method}

\begin{figure}[!htbp]
 \centering
\subfigure[]{
\begin{minipage}[t]{0.85\linewidth}
\centering
\includegraphics[width=1\columnwidth]{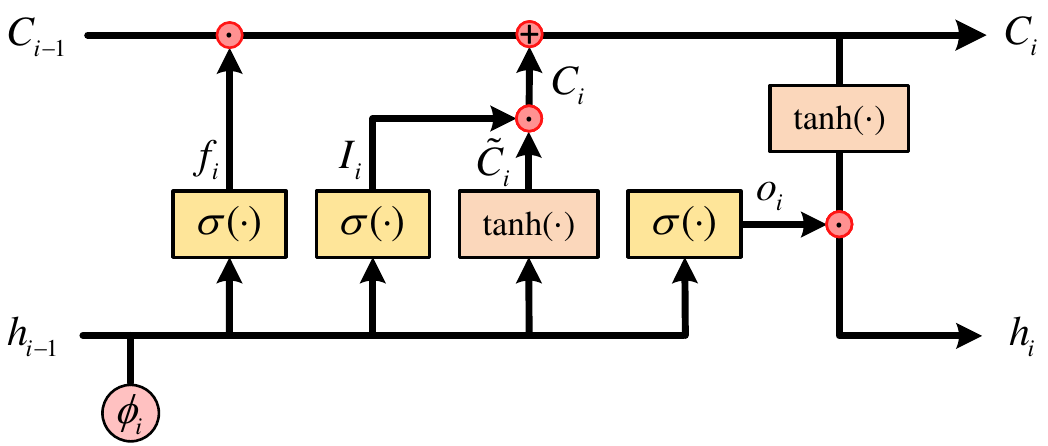}\label{lstm_}
\end{minipage}
}
\subfigure[]{
\begin{minipage}[t]{0.75\linewidth}
\centering
\includegraphics[width=1\columnwidth]{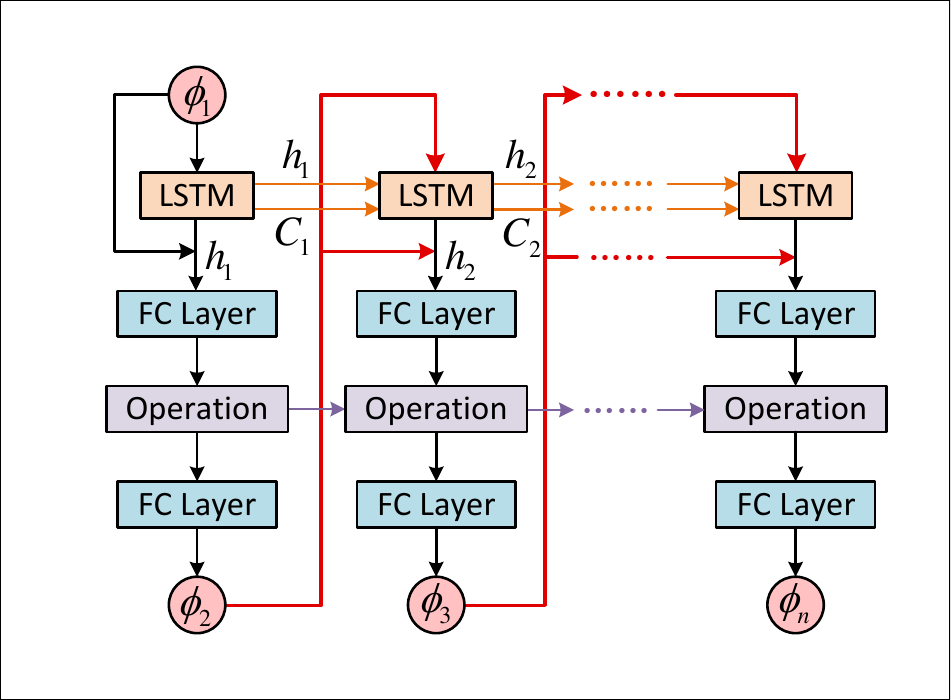}\label{cnp_lstm_}
\end{minipage}
}
\caption{\textbf{a)} the flowchart of the LSTM cell. The previous hidden state $h_{i-1}$ and current input $\phi_i$ are transformed to get $f_i$, $I_i$, ${\widetilde{C}}_{i}$, and $o_i$. The current cell state $C_i$ then can be obtained by adding the product of $f_i\cdot C_{i-1}$ and candidate value ${\widetilde{C}}_{i}$. The new hidden state $h_i$ is calculated by multiplying $\tanh (C_{i})$ and $o_{i}$. \textbf{b)} a general framework of neural network to predict order parameters. The order parameters ($\phi_i$, $i=1,$ $2,$ $\cdots,$ $m$) are processed by LSTM cells and get $r_i$. Then a commutative operation transform a combination of $r_1$, $r_2$, $\cdots$, $r_1$ to $R_i\in \mathcal{R}$. Then the predicted order parameter for the next time step $\phi_{i+1}$ ($j=m,$ $m+1,$ $\cdots,$ $n-1$) is obtained. }
\end{figure}

The following operations are applied to $\phi^p$ $(1 \leqslant p \leqslant N)$ independently, and we skip the superscript for convenience. Given a set of order parameters $\{\phi_i\}_{i=1}^{m}$ as the input, a NN is developed to predict the order parameters $\{\phi_i\}_{i=m+1}^n$. 

The order parameters are firstly encoded by an LSTM, whose workflow is shown in Figure \ref{lstm_}. LSTM is effective to solve long-term dependent problems because it introduces the forget gate $f_i$ in each LSTM cell, see Figure \ref{lstm_}, to control the circulation and loss of the sequential data. For each LSTM cell, the input of an LSTM cell includes the output of its previous cell, and all LSTM cells share the same structure and parameters. LSTM cells are then connected by the hidden states and cell states to form a complete LSTM network. In the following content, we will formulate the structure of an LSTM cell and then explain how LSTM cells connect with subsequent parts.
Inside the $i$ th LSTM cell $(1 \leqslant i \leqslant m-1)$, a \textcolor{black}{forget} gate $f_i \in [0,1]$ is defined as a Sigmoid unit \cite{olah2015understanding},
\begin{equation}\label{forget_gate}
 f_i = S (b_f + {U_f}\cdot \phi_{i} + {W_f}\cdot h_{i - 1}),
\end{equation}
where $S(\cdot)$ is the sigmoid function, $h_{i - 1}$ is the hidden state in the previous LSTM cell. Here, $b$, $U$, and $W$ are the bias, input weight, and cycle weight, respectively, and their subscript $f$ represents the forget gate. The same format is used in Eqs.(\ref{input_gate})-(\ref{output_gate}).

Then, an input gate $I_{i}$, i.e.,
\begin{equation}\label{input_gate}
I_{i} = S (b_i + {U_i} \cdot \phi_{i} + {W_i}\cdot h_{i - 1}),
\end{equation}
decides which information needs to be updated. 
To obtain the candidate value $\widetilde{C}_i$, a hypertangent activation function is applied,
\begin{equation}\label{candidate}
 {\widetilde{C}}_{i} = \tanh (b_{\widetilde{C}} + U_{\widetilde{C}} \cdot \phi_{i} + W_{\widetilde{C}} \cdot h_{i - 1}).
\end{equation}
Then, we obtain the current cell state $C_i$, incorporating the previous cell state $C_{i-1}$, the candidate value $\widetilde{C}_i$ in Eq. (\ref{candidate}), the \textcolor{black}{forget} gate $f_i$ in Eq. (\ref{forget_gate}), and the \textcolor{black}{input} gate $I_i$ in Eq. (\ref{input_gate}), i.e.,
\begin{equation}\label{cell_state}
 C_{i} = f_{i}\cdot C_{i-1}+I_{i}\cdot {\widetilde{C}}_{i}.
\end{equation}
The cell states can be regarded as the memory of LSTM, which connects the previous, current, and future LSTM cells. 
Next, we move to an output gate $o_i$, which decides the general state of the next hidden state. Here we use the sigmoid function as the activation function and project $\phi_i$ to the output gate $o_i$ by 
\begin{equation}\label{output_gate}
o_{i} = S (b_o + {U_o} \cdot \phi_{i} + {W_o} \cdot h_{i - 1}),
\end{equation}
where $h_{i-1}$ is the previous-step hidden state.
Finally, hidden state $h_i$ is computed from the current cell state $C_i$ in Eq. (\ref{cell_state}) and the output gate $o_i$ in Eq. (\ref{output_gate}) with
\begin{equation}
h_{i} = \tanh (C_{i})\cdot o_{i}.
\end{equation}

\textcolor{black}{The output range for sigmoid is $[0, 1]$, while tanh is $[-1, 1]$. The input gate, forget gate, and output gate use sigmoid because they control the quantities between 0 and 1, where 0 means no flow and 1 means complete flow. The estimated means of cell state $C_i$ and hidden state $h_i$ are expected to be 0, so we consider tanh function.} We have completed the operations in an LSTM cell. Once the hidden state $h_i$ is computed, we send current hidden state $h_i$ to a quasi encoder-decoder model. The representation $r_i$ is defined as a projection of hidden state $h_{i}$ and order parameter ${\phi_i}$:
\begin{equation}\label{cnplstm_encode}
 r_i = H_1\left( {{\phi_i},\ {h_{i}}} \right),
\end{equation}
where $H(\cdot)$ is an affine transformation using a non-linear activation (usually this is a sigmoid function):
\begin{equation}
 H_1\left( {{\phi_i},{h_{i}}} \right) =S \left( \left[ {{\phi_i}, \  {h_{i}}}\right]^T \cdot \boldsymbol {A_1}+ {b_1}\right).
\end{equation}
Here we first concatenate vectors $\phi_i$ and $h_{i}$ as a new vector, transpose the new vector, multiply it by the parameter matrix $\boldsymbol {A_1}$, and then add the offset parameter ${b_1}$. We will do the same affine transformation in Eq. (\ref{cnplstm_decode}) but with different parameters. Then a commutative operation maps all the representations $\{r_i\}_{i=1}^{m-1}$ into a one-dimensional element $R_i$ with the following rules:
\begin{equation}\label{cnplstm_hidden}
 R_i = r_1\oplus r_2\oplus \cdots \oplus r_i,
\end{equation}
where $\oplus$ here is the mean operation. Then we use another affine transformation with a non-linear activation to map the elements $R_i$ into $\widehat\phi_{i+1}$:
\begin{equation}\label{cnplstm_decode}
\widehat \phi_{i+1} = H_2\left(R_i\right),
\end{equation}

To update the parameters in the NN model, we defined a loss function to calculate the difference between the predicted $\phi$ and their observations. Given the observed order parameters $\phi_{i+1}$ and the prediction $\widehat\phi_{i+1}$, the loss function $\mathcal L_\phi(\boldsymbol\theta_1)$ is defined as:

\begin{equation}\label{lossfunc1}
\begin{array}{lll}
 \mathcal{L}_\phi(\boldsymbol \theta_1 ) = \displaystyle\frac{1}{N_C}\sum_{n_C}\left( \phi_{i+1}-\widehat\phi_{i+1} \right)^2,\\ 
 \end{array}
\end{equation}
where $N_C$ here is the number of predicted order parameters in the domain $\Omega$, and $n_C$ denotes the index of the grid cells. In practice, we minimize $\mathcal{L}_\phi(\boldsymbol \theta_1 )$ by Adam optimizer \cite{kingma2014adam} (see Algorithm \ref{adamupdate}).

\subsection{The multiphase consistent and conservative boundedness mapping algorithm}\label{conceive}


After predicting $\{\phi_k^p\}_{p=1}^N$ $(m+1 \leqslant k \leqslant n)$ from all the previous data before timestamp $k$ by a NN model in Section \ref{cnp_lstm_method}, there is no guarantee that $\{\phi_k^p\}_{p=1}^N$ satisfy the summation constraint, the mass conservation, and the boundedness constraint, which are the physical constraints on the order parameters, as discussed in Section \ref{Sec_Problem_definition}. Even though errors due to violating those constraints are small in a time step, they will accumulate and become significant as the computation proceeds. The multiphase consistent and conservative boundedness mapping (MCBOM) algorithm proposed in \cite{huang2021consistentVolume} is implemented to correct those small errors in each time step. The resulting $\{\phi_b^p\}_{p=1}^N$ not only satisfies the three physical constraints, but also is reduction-consistent with the given data $\{\phi_k^p\}_{p=1}^N$. This means that the absent locations of phase $p$, labeled by $\phi_b^p$, include those labeled by $\phi_k^p$ ($1 \leqslant p \leqslant N$). Then, $\{\phi_b^p\}_{p=1}^N$ are considered as the order parameters at the new time step. Since MCBOM is performed for every time-stamp, the subscript is omitted here.

Given a set of order parameters $\{\phi^p\}_{p=1}^N$ and a set of scalars $\{\Phi^p\}_{p=1}^N$ that represents total amounts of $\{\phi^p\}_{p=1}^N$ inside the domain of interest, the boundedness mapping includes three steps. The first one is the clipping step:
\begin{equation}\label{clipping}
\phi_{b*}^p=\left\{
    \begin{array}{ll}
    1,&\phi^p \geqslant 1, \\
    -1, &\phi^p \leqslant -1, \\
    \phi^p, &\mathrm{else}.
    \end{array}\right.
\end{equation}
The second one is the re-scaling step
\begin{equation}
    C^p_{b*}=\frac{1+\phi^p_{b*}}{2},
    \quad
    C^p_{b**}=\frac{C^p_{b*}}{\sum_{q=1}^N C^q_{b*}},
    \quad
    \phi^p_{b**}=2C^p_{b**}-1.
\end{equation}
The last step is the conservation step
\begin{equation}
    \phi^p_b=\phi^p_{b**}+\sum_{q=1}^N W^{p,q}_{b**} B^q,
\end{equation}
where $W^{p,q}$ is the weight function
\begin{equation}
    W^{p,q}=
    \left\{\begin{array}{ll}
    -(1+\phi^p)(1+\phi^q), &p \neq q,\\
    (1+\phi^p)(1-\phi^p), &p=q,
    \end{array}\right.
\end{equation}
and $\{B^p\}_{p=1}^N$ are obtained from solving the linear system
\begin{equation}\label{eq.mass}
    \left[ \int_{\Omega} W^{p,q}_{b**} d\Omega \right] [B^p]= \left[ \Phi^p-\int_{\Omega} \phi^p d\Omega \right].
\end{equation}
In practice, the integral is approximated by the mid-point rule. 

MCBOM corrects the prediction given by NN so that the properties of the order parameters are strictly satisfied. The resulting $\{\phi^p_b\}_{p=1}^N$ has the following properties:
\begin{equation}\label{conserve}
    \sum_{q=1}^N \frac{1+\phi^q_b}{2}=1,
    \quad
    |\phi^p_b| \leq 1,
    \quad
    \int_{\Omega} \phi^p_b d\Omega=\Phi^p,
    \quad
    \phi^p_b|_{\phi^p \leqslant -1},
    \quad
    1 \leqslant p \leqslant N.
\end{equation}

\textcolor{black}{The cost of the MCBOM algorithm is negligible compared to the neural networks because it is an pointwise algebraic operator. The major cost of MCBOM comes from inverting an $N$ by $N$ symmetric and diagonal-dominant matrix, where $N$ is the number of the phases that is usually much smaller than the parameter space of the NNs. Interested readers can refer to Huang \emph{et al.} \cite{huang2021consistentVolume} for more details.}

\subsection{The neural network with physical constraints}\label{pinns}

Once the order parameters are predicted from the first NN model in Section \ref{cnp_lstm_method} and corrected by MCBOM in Section \ref{conceive}, the density of the mixture is computed from Eq. (\ref{Eq Density}). Thanks to satisfying the boundedness constraint of the order parameters, the density of the fluid mixture will stay in their physical interval as well, no matter how large the density ratio of the problems could be. The remaining task is to predict the velocity $\{v_j\}_{j=m+1}^{n}$ given the observations $\{v_i\}_{i=1}^{m}$. This is accomplished by following the second NN model, whose structure is the same as the one in Section \ref{cnp_lstm_method}, but the input here becomes the velocity rather than the order parameters, i.e.,
\begin{equation}\label{cnplstm_v}
    \left\{\begin{array}{ll}
        f'_i = S (b'_f + {U'_f}\cdot v_{i} + {W'_f}\cdot h'_{i - 1}),\\[1.5mm]
        I'_{i} = S (b'_i + {U'_i} \cdot v_{i} + {W'_i}\cdot h'_{i - 1}),\\[1.5mm]
        {\widetilde{C}}'_{i} = \tanh (b'_{\widetilde{C}} + U'_{\widetilde{C}} \cdot v_{i} + W'_{\widetilde{C}} \cdot h'_{i - 1}),\\[1.5mm]
        C'_{i} = f'_{i}\cdot C'_{i-1}+I'_{i}\cdot {\widetilde{C}}'_{i},\\[1.5mm]
        o'_{i} = S (b'_o + {U'_o} \cdot v_{i} + {W'_o} \cdot h'_{i - 1}),\\[1.5mm]
        h'_{i} = \tanh (C'_{i})\cdot o'_{i},\\[1.5mm]
        r'_i = H'_1\left( {{v_i},{h'_{i}}} \right),\\[1.5mm]
        R'_i = r'_1\oplus r'_2\oplus \cdots \oplus r'_i,\\[1.5mm]
        \widehat v_{i+1} = H'_2\left(R'_i\right).\\
    \end{array}\right.
\end{equation}
Here $(\cdot)'$ denotes the variables and parameters in the second NN model for velocity prediction.

\begin{figure}[!htbp]
 \centering
  \includegraphics[width=.75\columnwidth]{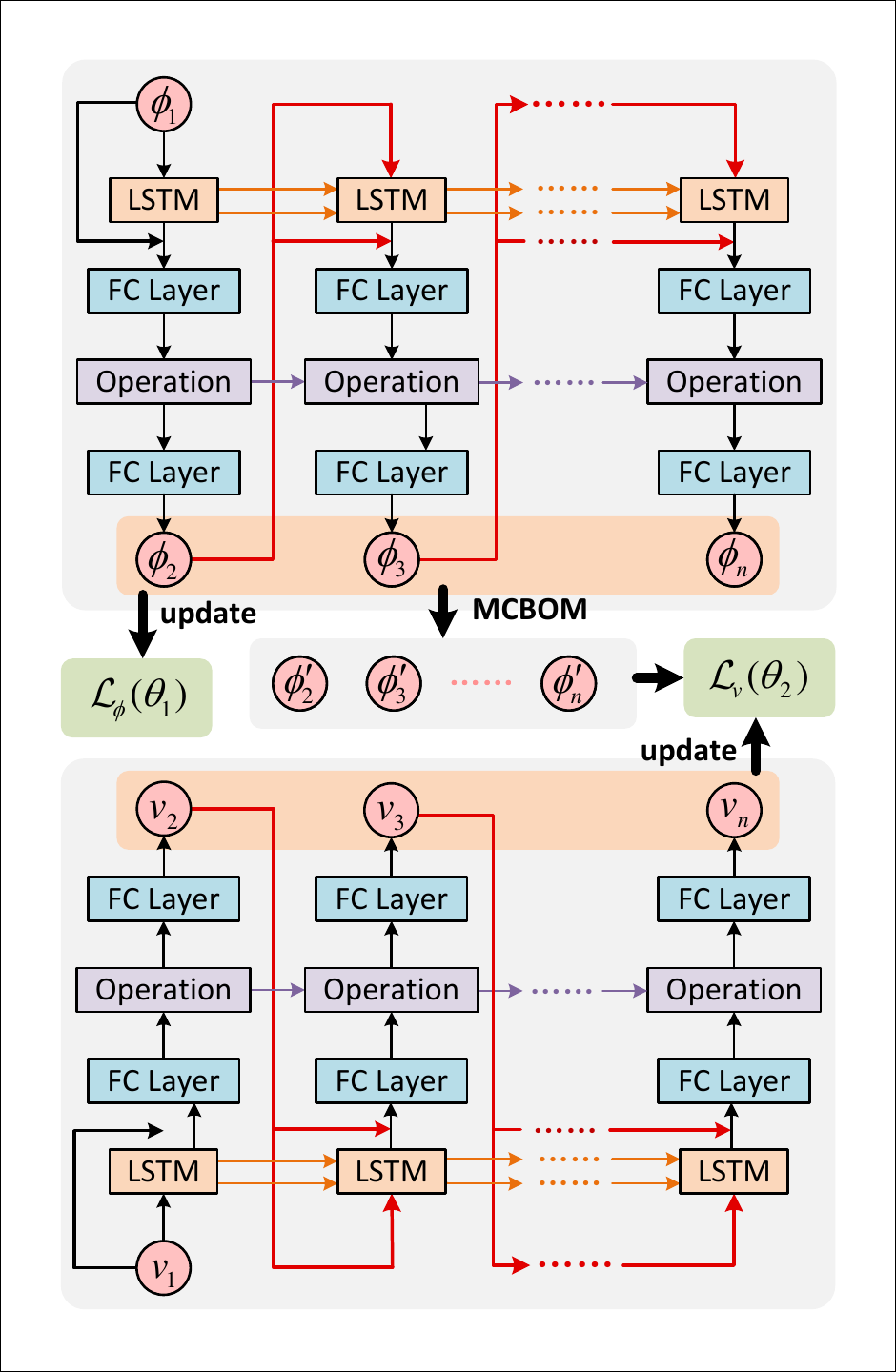}\\
 \caption{A general framework of the PCNN model. The observations $\phi_{i}$ are encoded in the first NN to predict $\widehat \phi_{i+1}$ at first. This prediction will be modified by the MCBOM algorithm and get $\phi'_{i+1}$. The modified $\phi_{i+1}$ will constrain the update of $v_{i+1}$ as a form of regularization term in the loss function of the second NN. $\widehat v_{i+1}$ will be predicted by the second NN.}\label{marc_lstm}
\end{figure}

The loss function of NN additionally considers the momentum conservation of the MPFs, which is different from the model in Section \ref{cnp_lstm_method}. The momentum of the multiphase flow is 
\begin{equation}\label{momentum}
  \mathcal M_i = \sum_{n_C}\left[\rho_i {v}_i \Delta \Omega \right]_{n_C},
\end{equation}
where $\mathcal M_i$ is the total momentum of the multiphase flow at timestamp $i$, $\Delta \Omega$ is the discrete volume. Recall that the density of the fluid mixture at timestamp $i$, i.e., $\widehat \rho_i$, is computed from Eq. (\ref{Eq Density}) once $\{\widehat \phi_i^p\}_{p=1}^N$ are available. Finally, the loss function of NN for velocity prediction is
\begin{equation}\label{lossfunc2}
 \mathcal{L}_v(\boldsymbol \theta_2 ) = \displaystyle\frac{1}{N_C}\sum_{n_C}\left( v_{i+1}-\widehat v_{i+1} \right)^2 +\lambda \left( {\sum\limits_{{n_C}} {{{\left[ {{{\widehat\rho }_{i + 1}}{{\widehat v}_{i + 1}}\Delta \Omega } \right]}_{{n_C}}}}  - {{\cal M}_{i + 1}}} \right)^2,
\end{equation}
where $\lambda$ is the Lagrange multiplier to enforce the momentum conservation. Here, ${\widehat\rho}_{i + 1}$ and ${\widehat v}_{i + 1}$ form a physics-informed constraint and penalize a nonphysical solution. Right now, the loss function in Eq. (\ref{lossfunc2}) not only includes the part to minimize the difference between the prediction and the ground truth, but also has the penalty due to violating the momentum conservation. A flowchart of the proposed PCNN in Sections \ref{cnp_lstm_method}-\ref{pinns} is shown in Figure~\ref{marc_lstm}. The methods to update the NN parameters and predictions are elaborated in Algorithm~\ref{adamupdate}. Including the physical constraints help to alleviate the nonphysical behaviors produced simply by using the traditional NNs, and those effects will then be shown in the next section.

{ \begin{algorithm}[H]\scriptsize
  \caption{Update strategy using Adam optimization \cite{kingma2014adam}.}
  \label{adamupdate}
   \begin{algorithmic}[1]
	\Require{Decay rate: $\rho_\phi$ and $\rho_v$, momentum coefficient $\alpha$.};
	\Require{Observations: order parameters \{$\boldsymbol\phi_{1}$, $\boldsymbol\phi_{2}$, $\cdots$, $\boldsymbol\phi_{m}$\} and  velocities \{$\boldsymbol v_{1}$, $\boldsymbol v_{2}$, $\cdots$, $\boldsymbol v_{m}$\}}.
	\Require{Learning rate for order parameters and velocities: $\eta$, $\eta'$.}
	\Require{Exponential decay rates: $\beta_1$, $\beta_2$.}
	\Require{Small constant for stabilization: $\epsilon$.}
	\State Initialize time stamp: $i=1$.
	\State Initialize the first and second moment: $\boldsymbol s=0$, $\boldsymbol r=0$, $\boldsymbol s'=0$, and $\boldsymbol r'=0$.
	\While{stopping criterion not met}
	\While{$i\neq n$}
	\Procedure{update $\widehat{f}_1$}{$\boldsymbol\phi_i, \boldsymbol\theta_1$}
	\State Update prediction: $\widehat{\boldsymbol \phi}_{i+1} \leftarrow \widehat{f}_1(\boldsymbol\phi_i, \boldsymbol\theta_1),$.
	\Comment{Eqs.(\ref{forget_gate}) - (\ref{cnplstm_decode})}.
	\State Update loss: $\mathcal{L}_\phi(\boldsymbol \theta_1 ) \leftarrow \displaystyle\frac{1}{N_C}\sum_{n_C}\left( \phi_{i+1}-\widehat\phi_{i+1} \right)^2$.
	\Comment{Eq. (\ref{lossfunc1})}
	\State Compute gradient: $\boldsymbol g\leftarrow \boldsymbol g-\nabla \mathcal{L}_\phi(\boldsymbol \theta_1)$.
	\State Update the first moment estimate: $\boldsymbol s\leftarrow\beta_1 \cdot \boldsymbol s+(1-\beta_1) \boldsymbol g$.
	\State Update the second moment estimate: $\boldsymbol r\leftarrow\beta_2 \cdot \boldsymbol r+(1-\beta_2) \boldsymbol g\odot \boldsymbol g$.\\
	\Comment{$\odot$ is the Hadamard product \cite{horn2012matrix}}.
	\State Correct the first moment estimation: $\hat{\boldsymbol s}\leftarrow \frac{\boldsymbol s}{1-\beta_1}$.
	\State Correct the second moment estimation: $\hat{\boldsymbol r}\leftarrow \frac{\boldsymbol r}{1-\beta_2}$.
	\State Update parameters: $\boldsymbol \theta \leftarrow \boldsymbol \theta - \eta \frac{\hat{\boldsymbol s}}{\sqrt{\hat{\boldsymbol r}}+\epsilon} $.
	\EndProcedure
	\State Modification: $\boldsymbol \phi_{i+1}' \leftarrow \widehat{\boldsymbol \phi}_{i+1}$.
	\Comment{Eqs.(\ref{clipping}) - (\ref{conserve})}.
	\State Transformation: $\widehat{\boldsymbol \rho}_{i+1} \leftarrow \boldsymbol \phi_{i+1}'$.
	\Comment{Eq. (\ref{Eq Density})}.
	\Procedure{update $\widehat{f}_2$}{$\boldsymbol v_i, \boldsymbol\theta_2$}:
	\State Update prediction: $\widehat{\boldsymbol v}_{i+1} \leftarrow \widehat{f}_2(\boldsymbol v_i, \boldsymbol\theta_2)$.\Comment{Eq. (\ref{cnplstm_v})}.
	\State Update loss: $ \mathcal{L}_v(\boldsymbol \theta_2 ) \leftarrow \displaystyle\frac{1}{N_C}\sum_{n_C}\left( v_{i+1}-\widehat v_{i+1} \right)^2 +\lambda \left( {\sum\limits_{{n_C}} {{{\left[ {{{\hat \rho }_{i + 1}}{{\hat v}_{i + 1}}\Delta \Omega } \right]}_{{n_C}}}}  - {{\cal M}_{i + 1}}} \right)^2$.\Comment{Eq. (\ref{lossfunc2})}.
	\State Compute gradient: $\boldsymbol g'\leftarrow \boldsymbol g' - \nabla\mathcal{L}_v(\boldsymbol \theta_2 )$.\\ 
	\State Update the first moment estimate: $\boldsymbol s'\leftarrow\beta_1 \cdot \boldsymbol s'+(1-\beta_1) \boldsymbol g'$.
	\State Update the second moment estimate: $\boldsymbol r'\leftarrow\beta_2 \cdot \boldsymbol r'+(1-\beta_2) \boldsymbol g'\odot \boldsymbol g'$.
	\State Correct the first moment estimation: $\hat{\boldsymbol s}'\leftarrow \frac{\boldsymbol s'}{1-\beta_1}$.
	\State Correct the second moment estimation: $\hat{\boldsymbol r}'\leftarrow \frac{\boldsymbol r'}{1-\beta_2}$.
	\State Update parameters: $\boldsymbol\theta' \leftarrow \boldsymbol \theta' - \eta' \frac{\hat{\boldsymbol s}'}{\sqrt{\hat{\boldsymbol r}'}+\epsilon} $.
	\EndProcedure
	\State Update time step: $i \leftarrow i+1$.
	\EndWhile\label{euclidendwhile}
	\EndWhile
	\State \textbf{return} \{$\boldsymbol \phi_{m+1}$, $\boldsymbol \phi_{m+2}$, $\cdots$, $\boldsymbol \phi_{n}$\} and \{$\boldsymbol v_{m+1}$, $\boldsymbol v_{m+2}$, $\cdots$, $\boldsymbol v_{n}$\}.
   \end{algorithmic}
\end{algorithm}
}

\section{Experiments}\label{experiment}

We will elaborate on two experiments here to demonstrate the effectiveness of the proposed PCNN. The first experiment is the horizontal shear layer problem and the other one is the dam break problem. The data sets are generated from the consistent and conservative phase-field method for incompressible MPFs \citep{huang2020consistentN}, including the boundedness mapping algorithm \citep{huang2021consistentVolume} that is also introduced in Section \ref{conceive}. The sequential data has been spatially discretized, and the discrete data set is summarized in Table \ref{dataset}, including the number of phases, simulation duration, number of time steps, and number of grids of the simulation. \textcolor{black}{The dataset used in the present study is generated by the physical multiphase flow model in Huang \emph{et al.} \cite{huang2021consistentVolume} which is a set of partial differential equations (PDEs), and the numerical methods therein. On the other hand, the present method is data-driven without solving any PDEs. The given data is on a uniform grid. The issue of using a random arrangement is that the total phasic mass and momentum in the entire domain, computed from Eqs. (\ref{eq.mass}) and (\ref{momentum}), are probably not accurate. If the given data on a random arrangement does not have that issue, the present model will still work.}

\begin{table}[htbp!]\centering
\caption{\textcolor{black}{A general summary of the data set.}}\label{dataset}
\begin{tabular}{ccccc}
\hline
\textbf{Dataset}       & \textbf{\textcolor{black}{\# phases}} & \textbf{\# time steps} & \textbf{Total time} & \textbf{Grid size}     \\ \hline
Horizontal shear layer & 3                    & 160                    & 2.0                 & {[}128 $\times$ 128{]} \\
Dam break              & 3                    & 200                    & 10.0                & {[}512 $\times$ 128{]} \\ \hline
\end{tabular}
\end{table}

\subsection{Horizontal shear layer simulation}\label{Sec HSL}

Horizontal shear flows happen under a variety of locations \cite{pereira2018challenges}, e.g., along coastal fronts, at the edges of energetic ocean currents, in oceans, and atmospheric currents around the terrain. Studies of the horizontal shear flows and predictions of their propagation trends can be widely applied to the hemodynamic analysis \cite{xu2017model}, ocean circulation \cite{christensen2018short}, etc.

In this case, the numerical domain of the problem is in a [1 $\times$ 1] box with the periodic boundary condition. The domain is partitioned into $128 \times 128$ non-overlapping, uniform cells. The time step is set to be $\Delta t$ = 0.0125. The density of the fluids here are: $\rho_1$ = 50.0, $\rho_2$ = 10.0 and $\rho_3$ = 1.0. At $t$=0, phase 1 occupies the region of $y\in(y_0,\ y_2]$ and keeps stationary. phase 2 is in between $y_1$ and $y_0$, and moves to the right with a unity speed. phase 3 is at the rest of the domain and moves to the left with a unity speed. Here we have $y_0$=0.5, $y_1$=0.25, and $y_2$=0.75. A sinusoidal vertical velocity is applied here to model the perturbation with an amplitude of 0.05 and wavelength of 2$\pi$. More details of the problem setup are available \cite{huang2020consistentN}.

\begin{figure}[!htbp]
\centering
\includegraphics[width=.95\columnwidth]{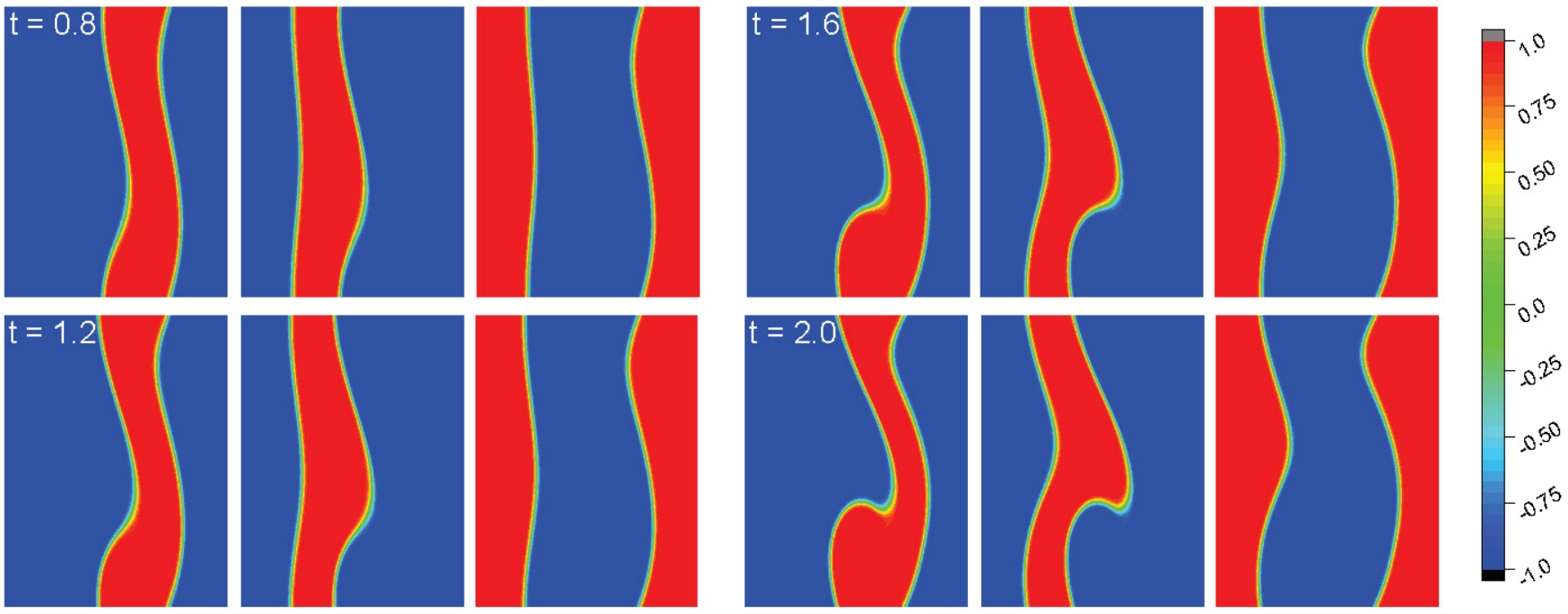}
\caption{Predicted order parameters of the three phases in the horizontal shear layer problem (given by NN-MCBOM). From top to bottom and left to right, $t=0.8,1.2,1.6,2.0$. Inside each panel, from left to right, values of the order parameters of phases 1, 2, and 3, respectively. Values of the order parameters indicate the locations of the phases. Where the order parameter of a phase is 1, only the corresponding phase occupies that location. Where the order parameter of a phase is -1, the corresponding phase is absent at that location. Here, we omit the comparison between the prediction results and the ground truth, because we cannot distinguish their differences according to the plots (the plots are almost the same).}\label{hslprediction}
\end{figure}

For the NNs, the encoder consists of a single feed-forward layer, recurrent layers with 64 LSTM cells, and two hidden layers with 32 neurons each. The activation function is ReLU. The learning rate is set to be 10$^{-4}$. The training set includes the data in the first 70\% of the considered time zone, while the rest 30\% is the validation set. \textcolor{black}{The neural network was trained on a desktop with Intel Core i9-10920X CPU, dual RTX 3080 GPUs, and 64GB DDR4 RAM memories. Training the neural network for simulating the horizontal shear layer takes 1 day.} The Part of the predictive results from the proposed PCNN model is shown in Figure \ref{hslprediction}.

\begin{figure}[!htbp]
\subfigure[]{
\centering
\includegraphics[width=.45\columnwidth]{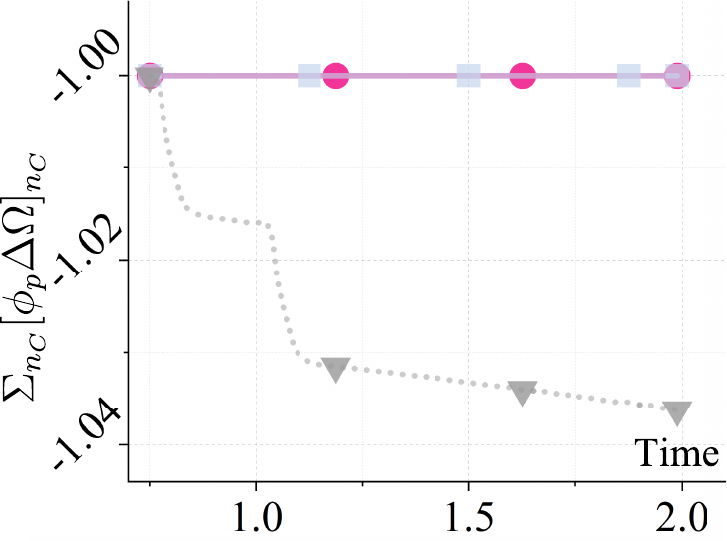}
\label{mass_conserve}
}
\subfigure[]{
\centering
\includegraphics[width=.45\columnwidth]{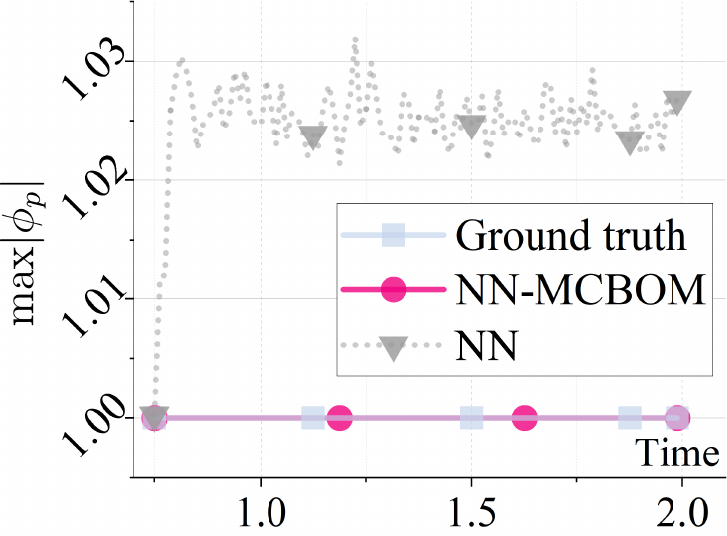}
\label{max_phase}
}

\subfigure[]{
\centering
\includegraphics[width=.45\columnwidth]{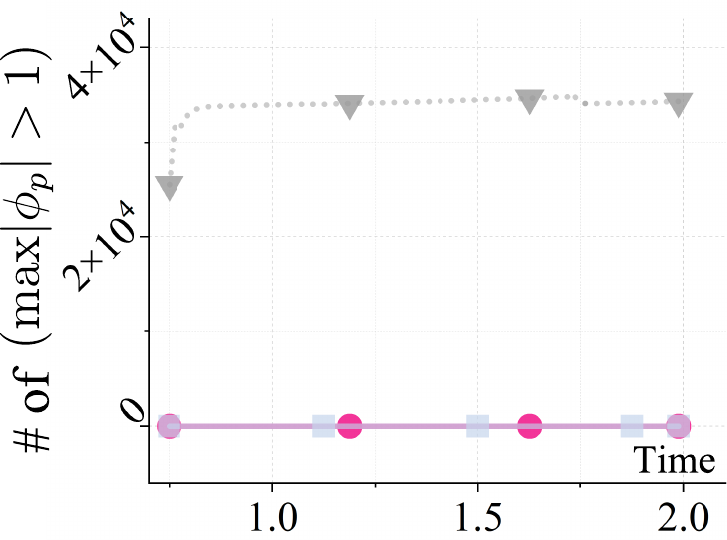}
\label{mass_bound}
}
\subfigure[]{
\centering
\includegraphics[width=.45\columnwidth]{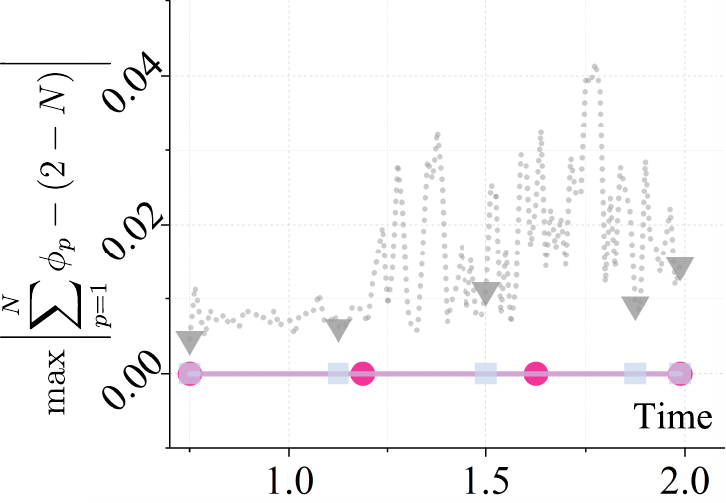}
\label{sum_phase}
}
\caption{a) ($\sum_{n_C} [\phi_1 \Delta \Omega]_{n_C}$) versus time. From the mass conservation, ($\sum_{n_C} [\phi_1]_{n_C}$) is independent of time. b) $\max |\phi_1|$ versus time. The order parameters should always fall in [-1, 1] or $\max |\phi_1|$ $\leq$ 1. c) Number of grid points where $\phi_1$ exceeds $[-1,1]$ versus time. d) $\max |\sum_{p=1}^N \phi_p-(2-N)|$ versus time. To avoid local voids or overfilling, the summation of the order parameters should always be $(2-N)$.}\label{all_pahse}
\end{figure}

We first illustrate the necessity of implementing MCBOM in Section \ref{conceive}, after obtaining predictions from the first NN model in Section \ref{cnp_lstm_method}. For a clear presentation, we only show the results of the first phase (phase 1) in Figure \ref{all_pahse}. For comparison, we present the results from the first NN model in Section \ref{cnp_lstm_method}, and the training loss converges, which, in our experiment, needs about 400 epochs. Figure \ref{mass_conserve} shows the mass change of phase 1 in the entire domain versus time. \textcolor{black}{The ground truths of the results are obtained from the numerical solver in Huang \emph{et al.} \cite{huang2021consistentVolume}. Therefore, the results we show are comparing to those with the method in Huang \emph{et al.} \cite{huang2021consistentVolume}.} It is clear that the mass of the phase is time-independent, after applying MCBOM, while the mass is fluctuating from the output of the NN model in Section \ref{cnp_lstm_method}. In other words, mass gain or loss appears in the neural network prediction, which violates the principle of mass conservation. This issue is successfully addressed with the help of MCBOM.
Figure \ref{max_phase} indicates the time history of the maximum absolute value of the phase. The order parameter is in between $-1$ and $1$, as mentioned in Section \ref{Sec_Problem_definition}, so that the density of the fluid mixture computed from the order parameters is also in its physical interval. Without performing MCBOM, the prediction from the NN model exceeds the expected interval, $[-1,1]$ even after a long time of training. This out-of-bound issue does not appear in the results of the proposed model, including MCBOM. 
Further, we count the number of grid points where the order parameter of the phase is beyond $[-1,1]$, or $\mathrm{max}|\phi_p|>1$, and the results are shown in Figure \ref{mass_bound}. As time goes on, more and more grid points have out-of-bound predictions from the NN model, and the number finally reaches around 34,000 here. With MCBOM, the out-of-bound issue is not observed, as expected.
Finally, we check the summation of the order parameters, which is related to the production of local void or overfilling, and the results are shown in Figure \ref{sum_phase}. As mentioned in Section \ref{Sec_Problem_definition}, the sum of the order parameters should always be $(2-N)$, which is $-1$ in the present case. However, the prediction from the NN model does not have this property because it is not enforced in the NN. The summation of the order parameters predicted from NN is observed oscillating between $0$ and $0.04$. On the other hand, the summation constraint of the order parameters is strictly satisfied with the help of MCBOM.

\begin{figure}[!htbp]
\centering
\includegraphics[width=.48\columnwidth]{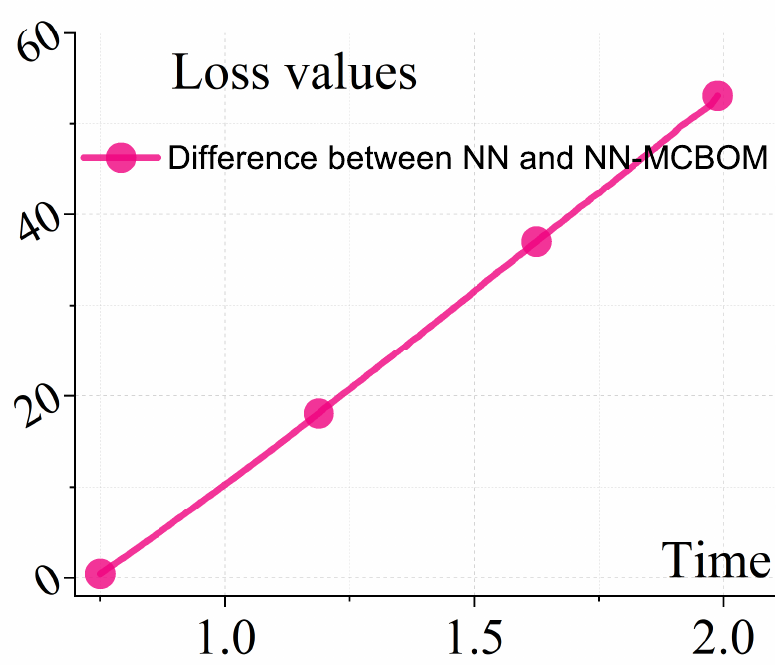}
\caption{\textcolor{black}{The total absolute error difference each time step. The ground truth of the problem is obtained with $[256 \times 256]$ grid points. The loss value at a certain time $t$ is obtained by adding the sum of MAEs of all spatial positions at time $t$. The difference is obtained by subtracting the total absolute error of the NN model from the one of the NN-MCBOM model. As time goes on, the difference in the errors increases and is always positive. Therefore, the NN-MCBOM prediction is more accurate and reliable.}}\label{mse_difference}
\end{figure}

\textcolor{black}{To indicate the improvement of accuracy, Figure \ref{mse_difference} compares the mean absolute errors (MAEs) between the prediction from the NN model or NN-MCBOM model and the ground truth of the problem obtained with $[256\times 256]$ grid points. For a clear presentation, the error from the NN model minus the one from the NN-MCBOM model versus time is shown. It should be noted that the difference in the errors is always positive, which indicates that the NN-MCBOM prediction is always closer to the ground truth than the one from the NN model and therefore more accurate. Moreover, the difference in the error is increasing with time, which indicates that the NN model only is less trustful for long time prediction.} Therefore, implementing MCBOM not only exactly enforces the physical constraints for the order parameters, but also produces more accurate and reliable long-time predictions.

\begin{figure}[!htbp]
\subfigure[]{
\centering
\includegraphics[width=.45\columnwidth]{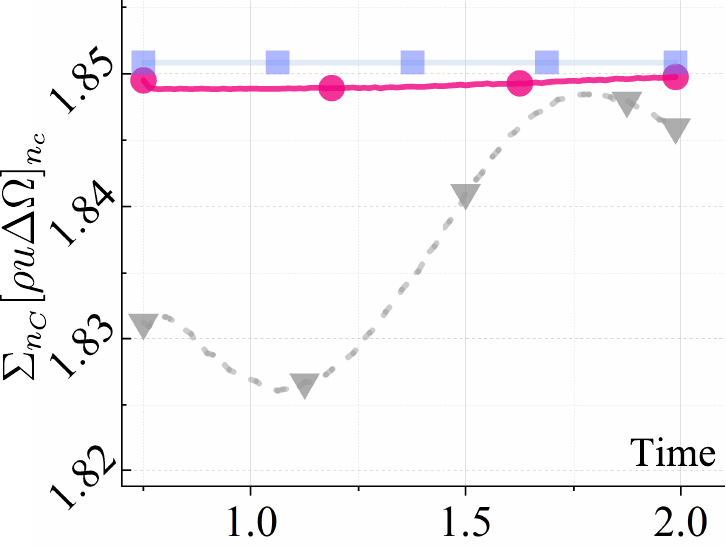}\label{momentum_x1}
}
\subfigure[]{
\centering
\includegraphics[width=.45\columnwidth]{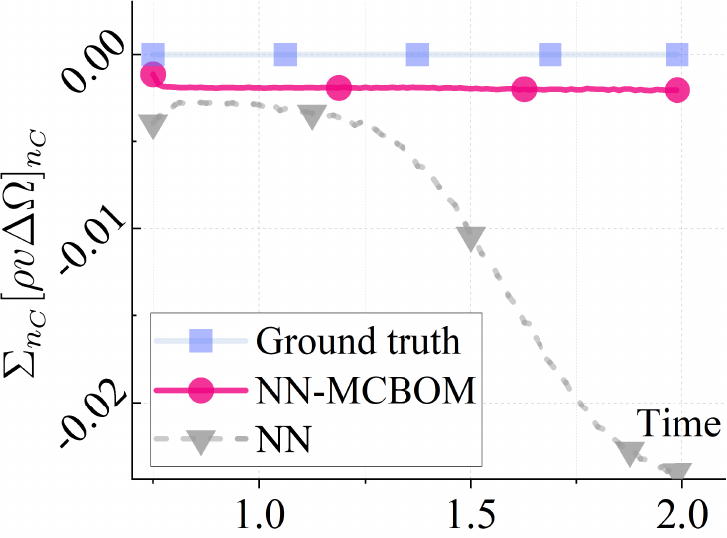}\label{momentum_y}
}
\caption{\textcolor{black}{a) $\sum_{n_C}[\rho u \Delta \Omega]_{n_C}$ versus time in the horizontal shear layer problem. b) $\sum_{n_C}[\rho v\Delta \Omega]_{n_C}$ versus time. The momentum in the problem is conserved, which is well predicted by the proposed PINN as compared to the ground truth, but the traditional NN model fails to predict this behavior.}}
\end{figure}

Figs. \ref{momentum_x1} and \ref{momentum_y} show the momentum in the $X$ and $Y$ directions predicted by the proposed NN in Section \ref{pinns} or the traditional NN model without including the momentum conservation in the cost function. The momentum predicted from the NN is almost independent of time, and its value is very close to the ground truth. On the other hand, the behavior predicted from the traditional NN is very different from the ground truth. We can conclude that the NN considered momentum conservation can provide a more physically plausible prediction.

\subsection{Dam break simulation}
\begin{figure}[htbp]
\centering
\includegraphics[width=.9\columnwidth]{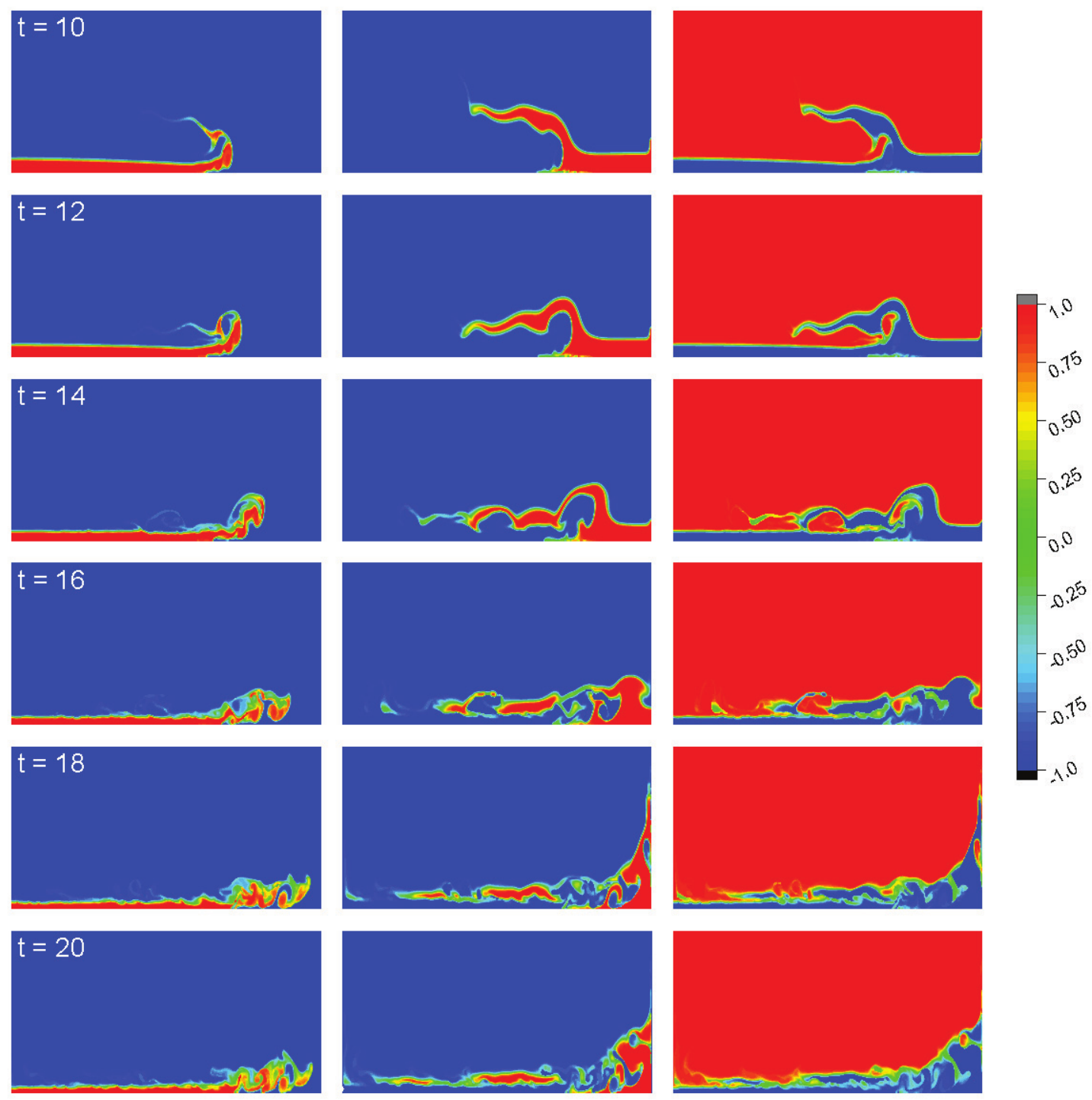}
\caption{Predicted order parameters of the three phases in the dam break problem (given by NN-MCBOM). From top to bottom, $t=10,12,...,20$. From left to right, values of the order parameters of phases 1, 2, and 3, respectively. Values of the order parameters indicate the locations of the phases. Where the order parameter of a phase is 1 (red), only the corresponding phase occupies that location. Where the order parameter of a phase is -1 (blue), the corresponding phase is absent at that location. Same as the first experiment, we omit the comparison between the ground truth and the prediction results.}\label{damprediction2}
\end{figure}

Extreme weather associated with rainfall often threatens dam safety. Dam break problem, as a low-frequency but high-hazard catastrophic factor that may cause significant losses, has received increasing attention and is widely studied. In recent years, the development of computer simulation techniques makes it possible of modeling the dynamics of the problem \cite{huang2020consistentN}. Also, dam break simulation is a good case of large-density-ratio problems. The first phase is considered as water, whose density is $\rho_1=829$. The second phase is the oil with the density of $\rho_2=463$. The third phase is considered air, whose density is set to be $\rho_3=1$. Initially, a water square is at the leftmost of the domain, while an oiled square is at the rightmost, and the fluids are stationary. The domain of the problem is [4 $\times$ 1] with the no-slip boundary condition. The number of grid points is [512$\times$128]. The time step is $\Delta t = 5 \times$ 10$^{-2}$. \textcolor{black}{We used the same environment and equipment as in the HSL simulation, which took 5 days to train the neural network.} The flows is  at $t=0$. More details of the problem setup are available \cite{huang2020consistentN}.

The PCNN here is identical to the one in the Section \ref{Sec HSL} except that it has 128 LSTM cells. Part of the predictive results is shown in Figure \ref{damprediction2}.

\begin{figure}[!htbp]
\subfigure[]{
\centering
\includegraphics[width=.45\columnwidth]{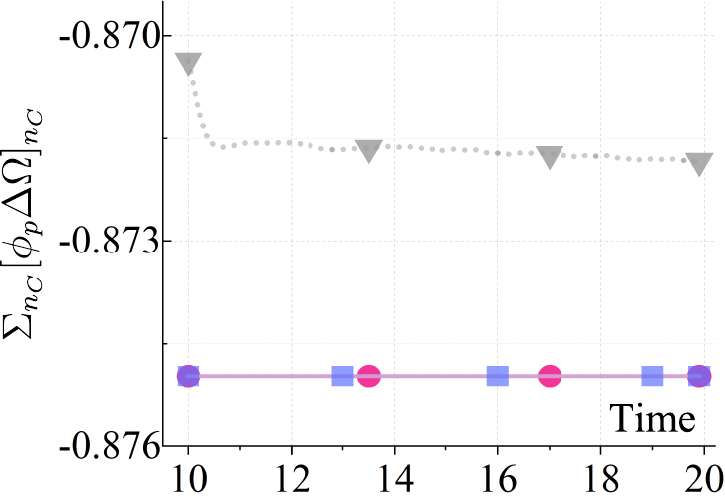}
\label{mass_conserve2}
}
\subfigure[]{
\centering
\includegraphics[width=.45\columnwidth]{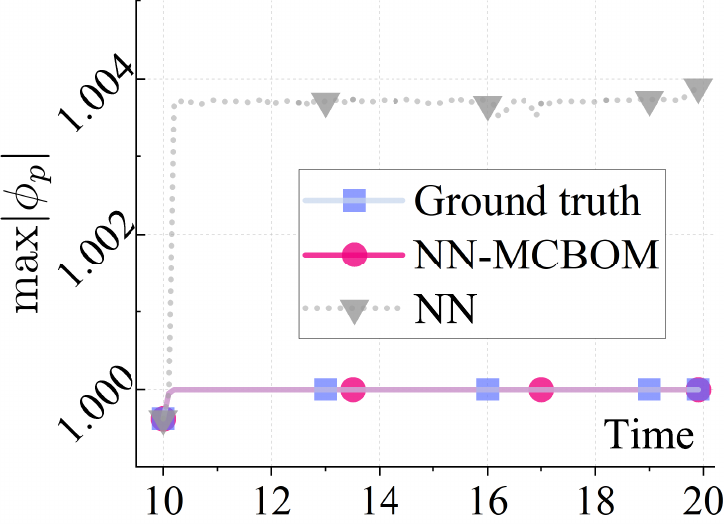}
\label{max_phase2}
}

\subfigure[]{
\centering
\includegraphics[width=.45\columnwidth]{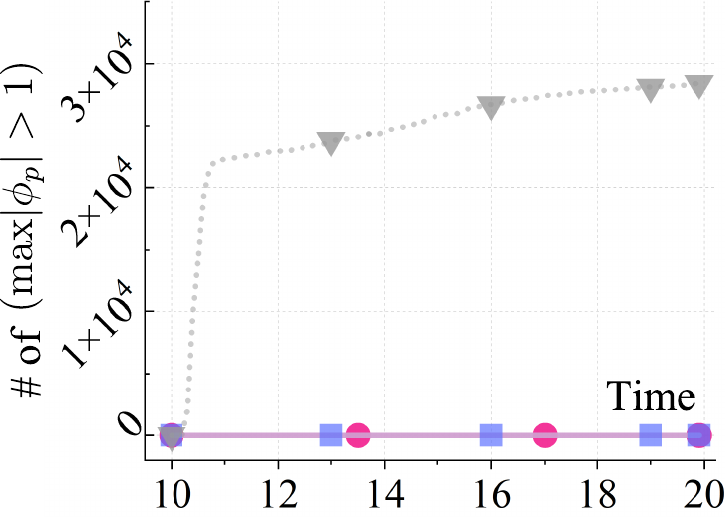}
\label{mass_bound2}
}
\subfigure[]{
\centering
\includegraphics[width=.45\columnwidth]{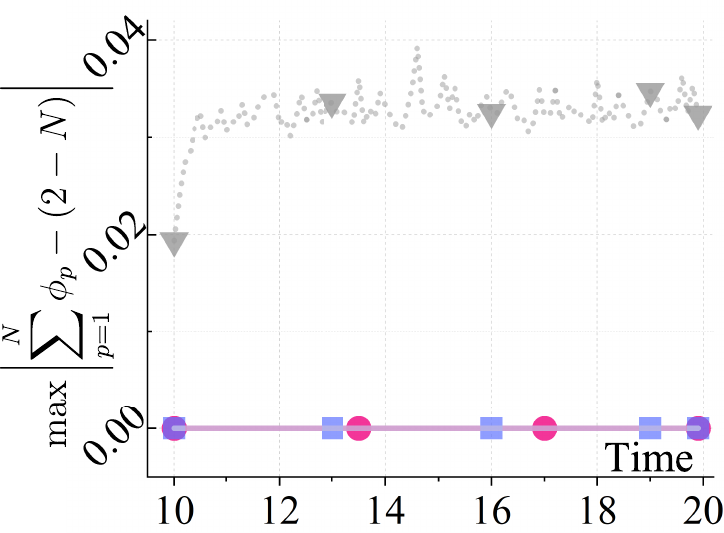}
\label{sum_phase2}
}
\caption{a) ($\sum_{n_C} [\phi_1 \Delta \Omega]_{n_C}$) versus time.  ($\sum_{n_C} [\phi_1 \Delta \Omega]_{n_C}$) should be independent of time due to mass conservation. b) $\max |\phi_1|$ versus time. The order parameters should fall in [-1, 1] or $\max |\phi_p|$ $\leq$ 1. c) Number of grid points where the order parameter exceeds $[-1,1]$ versus time; d) ($\mathrm{max}|\sum_{p=1}^N \phi_p-(2-N)|$) versus time. The summation of the order parameters should be $(2-N)$ so that no voids or overfilling is produced.}
\end{figure}

\begin{figure}[!htbp]
 \centering
  \includegraphics[width=.58\columnwidth]{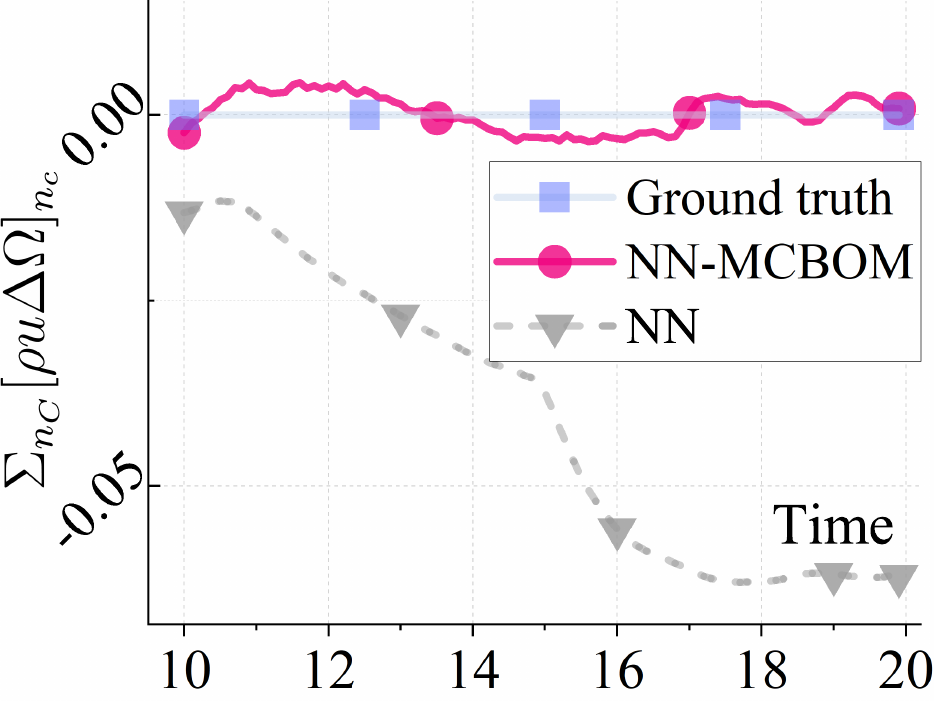}\\
 \caption{$\sum_{n_C}[\rho u \Delta \Omega]_{n_C}$ versus time in the dam break problem.}\label{momentum_x2}
\end{figure}

Figure \ref{mass_conserve2} shows the mass of the first phase (phase 1) versus time. According to the conservation of mass, the total amount of $\phi_1$ in the entire domain should not change. This is true in the prediction given by the NN-MCBOM, but the prediction given by the NN model violates this principle.
Figure \ref{max_phase2} shows the maximum absolute value of the order parameter of the first phase predicted by the proposed NN-MCBOM model or the NN model. It is obvious that the order parameter predicted by the NN model exceeds 1, which violates the boundedness constraints for the order parameters. This further indicates the necessity of applying MCBOM.
As shown in Figure \ref{mass_bound2}, there are about 28,000 grid points having an out-of-bound order parameter if only NN is used for prediction.
Figure \ref{sum_phase2} shows the summation of the order parameters. The proposed model strictly follows the summation constraint for the order parameters, while this is not the case without the help of MCBOM. The results given by the NN model oscillate between $0.02$ and $0.04$.
Figs. \ref{momentum_x2} shows the prediction of the momentum in the $x$ direction. Results predicted by the NN model will drop from $-0.01$ to $-0.07$. On the other hand, the prediction of the proposed NN including a physics-informed penalization of momentum conservation, is much closer to the validation data.

\section{Conclusion}\label{conclusion}

In the present study, a PCNN model is developed to predict MPFs. It consists of the first NN model to predict the order parameters, the MCBOM to correct the order parameters predicted by the NN model in order to exactly satisfy the mass conservation and the summation and boundedness constraints for the order parameters, and the second NN model to predict the flow velocity with a penalization of the momentum conservation using the density of the fluid mixture computed from the corrected order parameters.
We demonstrate that the prediction of the order parameters using only the NN model violates all the physical constraints. Hence, nonphysical behaviors, such as mass loss, local voids, or overfilling, are observed in our numerical experiments. After combining the NN and MCBOM, all those issues are thoroughly removed. Moreover, the accuracy and reliability of long-time predictions are improved when including the MCBOM. The momentum conservation is acceptably reproduced, although not exactly, by the proposed NN with physical constraints to predict the velocity. However, the traditional NN model fails to predict such a physical behavior.

In summary, we consider the NN model as the basic structure of the PCNN for MPFs, and enforce the physical constraints either implicitly by penalization or explicitly (or exactly) by MCBOM. The proposed PCNN model is an effective tool to predict the dynamics of the MPFs.

\section*{Acknowledgement}
The authors gratefully acknowledge the support of the National Science Foundation (DMS-1555072, DMS-2053746, and DMS-2134209), Brookhaven National Laboratory Subcontract 382247, and the U.S. Department of Energy (DOE) Office of Science Advanced Scientific Computing Research program DE-SC0021142).

\section*{Author Declarations}
\subsection*{Conflict of interest}
The authors have no conflicts to disclose.
\subsection*{Data availability}
The data that support the findings of this study are available
from the corresponding author upon reasonable request.

\bibliography{refs}

\end{document}